\documentclass{ar1col}
\usepackage{mathtools}
\usepackage{cite}

\jname{Submitted version. Posted with permission
from the Annual Review of Condensed Matter Physics}
\jyear{2017}

\begin{document}

\markboth{Hallas \textbullet~Gaudet \textbullet~Gaulin}{Quantum XY Pyrochlores}

\title{Experimental Insights into Ground State Selection of Quantum XY Pyrochlores}

\author{Alannah M. Hallas,$^1$ Jonathan Gaudet,$^1$ and Bruce D. Gaulin$^{1,2,3}$
\affil{$^1$Department of Physics and Astronomy, McMaster University, Hamilton, ON, L8S 4M1, Canada; email: hallasa@mcmaster.ca, gaudej@mcmaster.ca, gaulin@mcmaster.ca} 
\affil{$^2$Canadian Institute for Advanced Research, 180 Dundas St. W., Toronto, ON, M5G 1Z7, Canada}
\affil{$^3$Brockhouse Institute for Materials Research, McMaster University, Hamilton, ON L8S 4M1 Canada}}

\begin{abstract}
Extensive experimental investigations of the magnetic structures and excitations in the XY pyrochlores have been carried out over the last decade. Three families of XY pyrochlores have emerged: Yb$_2B_2$O$_7$, Er$_2B_2$O$_7$, and, most recently, $AA'$Co$_2$F$_7$. In each case, the magnetic cation (either Yb, Er, or Co) exhibits XY anisotropy within the local pyrochlore coordinates, a consequence of crystal field effects. Materials in these families display rich phase behavior and are candidates for exotic ground states, such as quantum spin ice, and exotic ground state selection via order-by-disorder mechanisms. In this review, we present an experimental summary of the ground state properties of the XY pyrochlores, including evidence that they are strongly influenced by phase competition. We empirically demonstrate the signatures for phase competition in a frustrated magnet: multiple heat capacity anomalies, suppressed $T_N$ or $T_C$, sample and pressure dependent ground states, and unconventional spin dynamics.
\end{abstract}

\maketitle

\section{INTRODUCTION}

\begin{figure}[bp]
\includegraphics[width=5in]{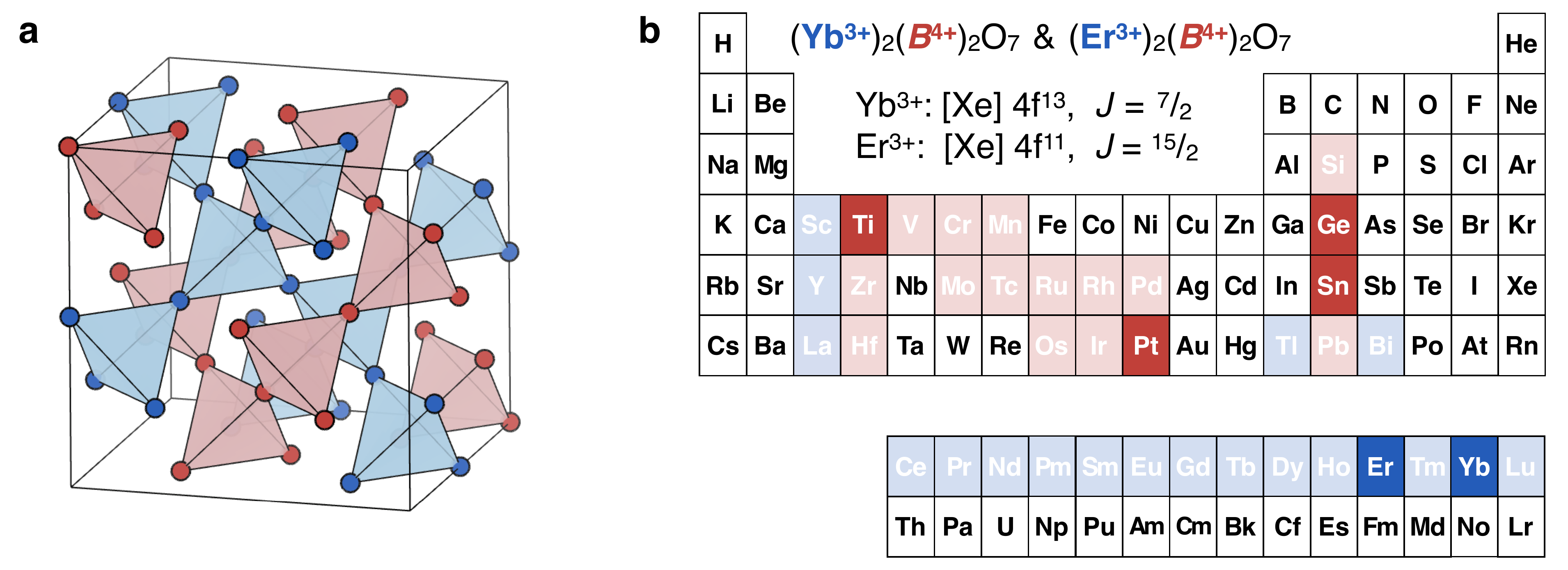}
\caption{(a) The $A$ (blue) and $B$ (red) sites of the pyrochlore lattice, $A_2B_2$O$_7$, each form a network of corner-sharing tetrahedra, a motif prone to magnetic frustration. (b) XY anisotropy on the pyrochlore lattice is achieved when the $A$ site is occupied by either erbium or ytterbium, as indicated by the blue highlighting, with the possible non-magnetic $B$ sites highlighted in red. The faded blue and red squares reflect all known $A^{3+}$ and $B^{4+}$ cations that can be found on the pyrochlore lattice in some combination.}
\label{PeriodicTable}
\end{figure}

Cubic pyrochlore oxides of the form $A_2B_2$O$_7$ display a broad range of physical properties, in direct relation to their chemical diversity. The $A$ and $B$ sites of this lattice can each host many chemical elements, leading to families that involve over 150 compounds. Pyrochlores can exhibit insulating, metallic, and even superconducting states \cite{subramanian1983oxide}. The experimental intrigue with pyrochlore magnetism began in earnest with the discovery of a spin glass state in Y$_2$Mo$_2$O$_7$, unexpected in a structurally well-ordered material \cite{greedan1986spin}. From that point on, pyrochlore oxides have been of great interest to physicists and at the forefront of new magnetic and electronic phenomena, including: cooperative paramagnetism in Tb$_2$Ti$_2$O$_7$~\cite{gardner1999cooperative}, the anomalous Hall effect in Nd$_2$Mo$_2$O$_7$~\cite{katsufuji2000anomalous,taguchi2001spin}, giant magnetoresistance in Tl$_2$Mn$_2$O$_7$~\cite{shimakawa1996giant}, and, most recently, possible topological states in $A_2$Ir$_2$O$_7$~\cite{wan2011topological}. Insulating rare earth pyrochlores in particular have attracted enormous attention as they represent an archetype for 3-dimensional geometrically frustrated magnetism. The $A$ and $B$ sites of the pyrochlore lattice each independently form a network of corner-sharing tetrahedra, as drawn in \textbf{Figure \ref{PeriodicTable}(a)}, an architecture that is prone to frustration when occupied by a magnetic cation. The whole series of rare earth cations, most of which are magnetic, can reside on the $A$ site of the lattice. These rare earth pyrochlores have been lauded for their diverse magnetic properties, which can be attributed to the combination of varied magnetic anisotropies and interactions that conspire to produce a wealth of exotic ground states~\cite{gardner2010magnetic}.

\begin{marginnote}[]
\entry{Frustration}{The inability of a system to satisfy all of its pairwise constraints simultaneously.}
\end{marginnote}

The rare earth pyrochlores were propelled into the spotlight when the classical spin ice state was discovered in Ho$_2$Ti$_2$O$_7$~\cite{harris1997geometrical,bramwell2001spin}, and soon after in Dy$_2$Ti$_2$O$_7$~\cite{ramirez1999zero}. The spin ice state has a residual entropy analogous to that of proton disorder in water ice~\cite{pauling1935structure}, which in both cases is the result of an extensive ground state degeneracy. The requirements for spin ice physics are the connectivity of the pyrochlore lattice, local Ising anisotropy, and net ferromagnetic interactions~\cite{harris1997geometrical}. The dipolar interaction between Ising moments on the pyrochlore lattice is ferromagnetic at the nearest neighbor level~\cite{den2000dipolar,isakov2005spin}. Consequently, provided antiferromagnetic exchange interactions are sufficiently weak, large Ising moments ($\sim10~\mu_B$) on the pyrochlore lattice, in and of themselves, have the ingredients for spin ice physics. Ten years after its discovery, the spin ice state was understood to possess emergent magnetic monopole excitations~\cite{castelnovo2008magnetic,morris2009dirac,fennell2009magnetic}, generating a new level of interest. However, the spin ice states found in Ho$_2$Ti$_2$O$_7$ and Dy$_2$Ti$_2$O$_7$ are predominantly governed by classical physics, in part because their crystal field ground state doublets are almost purely made up of maximal $m_J$~\cite{rosenkranz2000crystal,rau2015magnitude,ruminy2016crystal}. Thus, a route to enhancing quantum effects is to consider rare earths where contributions from smaller $m_J$ states are non-negligible, thereby inducing XY rather than Ising anisotropy. This is achieved in nature when the rare earth site is occupied by either ytterbium or erbium (\textbf{Figure \ref{PeriodicTable}(b)}), the subjects of this review.

The incorporation of XY degrees of freedom has important implications for the spin ice state. In the classical spin ices, which have large Ising moments, the magnetic interactions are dipolar augmented by relatively weak exchange~\cite{den2000dipolar,isakov2005spin}. If the magnitude of the rare earth moment is reduced, dipolar interactions become less important and exchange interactions, correspondingly, more important. Then, anisotropic exchange terms can couple effective $S=\frac{1}{2}$ moments, which will promote quantum fluctuations, particularly in the case of a ground state composed of minimal $m_J$. A possible resulting state is referred to as quantum spin ice. One of the most exciting revelations has been that the quantum spin ice problem maps onto an emergent quantum electrodynamics, whose elementary excitations include not only magnetic monopoles, but also electric monopoles and gauge photons~\cite{hermele2004pyrochlore,banerjee2008unusual,shannon2012quantum,savary2012coulombic,benton2012seeing,lee2012generic,savary2013spin,gingras2014quantum}. 
\begin{marginnote}[]
\entry{Effective $S=1/2$}{In the case of an isolated ground state doublet, the system can be projected into a pseudo-spin $\frac{1}{2}$ basis.}
\end{marginnote}In this realm, theory is out-pacing experiment, for a robust experimental realization of quantum spin ice remains elusive. Amongst the rare earth pyrochlores, several candidate materials have been identified as possessing many of the requisite ingredients, including: Tb$_2$Ti$_2$O$_7$~\cite{molavian2007dynamically}, Yb$_2$Ti$_2$O$_7$~\cite{ross2011quantum}, and Pr$_2$Zr$_2$O$_7$~\cite{kimura2013quantum}. While quantum spin ice ground states in these materials remain tantalizingly close, recent experimental studies show rich low temperature phenomenology including strong sensitivity to quenched impurities~\cite{taniguchi2013long,arpino2017impact,wen2016disordered}, which complicates a simple categorization of their ground states.

Another example of quantum XY phenomenology is found in the degeneracy-breaking that results in long range magnetic order in Er$_2$Ti$_2$O$_7$. In the mean field limit of its spin Hamiltonian, there is a degeneracy between two ordered, non-collinear XY states, termed $\psi_2$ and $\psi_3$. Linear combinations of these two states continuously span the XY plane, giving a U(1) degeneracy. Thus, it was at first perplexing when Er$_2$Ti$_2$O$_7$ was observed to undergo a continuous phase transition into a pure $\psi_2$ state~\cite{champion2003er,poole2007magnetic}. The selection of this ground state is believed to be driven by thermal and quantum fluctuations, i.e. order-by-disorder. Despite the mean field U(1) degeneracy, the spin excitations of $\psi_2$ and $\psi_3$ contribute differently to the free energy. This can occur either at finite temperature via a thermal order-by-disorder mechanism~\cite{oitmaa2013phase,zhitomirsky2014nature,javanparast2015order,yan2017theory}, or via quantum fluctuations at zero temperature, hence quantum order-by-disorder~\cite{savary2012order,zhitomirsky2012quantum,wong2013ground}. While theoretical ideas on order-by-disorder date back more than 60 years~\cite{tessman1954magnetic,villain1980order}, $\psi_2$ ground state selection in Er$_2$Ti$_2$O$_7$ remains one of the most compelling candidates for its experimental realization. However, an alternative mechanism that relies only on energetic selection by virtual crystal field transitions has also been shown to break the U(1) degeneracy in Er$_2$Ti$_2$O$_7$~\cite{mcclarty2009energetic,petit2014order,rau2016order}. 

This introduction has highlighted several fascinating aspects of the effective $S=\frac{1}{2}$ quantum XY pyrochlores, such as ground state selection by order-by-disorder and quantum spin ice. Here, we will focus our attention on the prevalence for low temperature magnetism that appears to be influenced by the presence and proximity of competing phases. We compare and contrast the properties of two XY families, Er$_2B_2$O$_7$ and Yb$_2B_2$O$_7$, with varying non-magnetic $B$ site cations, and conclude that many of their exotic properties stem from strong phase competition, in agreement with recent theoretical suggestions~\cite{yan2013living,jaubert2015multiphase,yan2017theory}.

\section{ORIGIN OF XY ANISOTROPY}

A unique attribute of rare earth magnetism is the hierarchy of energy scales, where spin-orbit coupling, the crystal electric field, and exchange interactions each have characteristic energies that are typically separated by an order of magnitude or more. The spin anisotropy of the rare earth pyrochlores is derived from crystal electric field effects. Thus, we begin by elucidating the origin of XY anisotropy and justify the application of the XY label to the ytterbium and erbium pyrochlores. As a consequence of their seven $f$ electron orbitals, magnetic rare earths typically have large values for their total angular momentum $J$, as compared to magnetic elements elsewhere on the periodic table. Furthermore, as rare earth ions are relatively heavy, the spin-orbit coupling, $\lambda$, is usually sufficiently large to isolate a spin-orbit ground state that is comprised of $2J+1$ degenerate levels. Within a crystalline environment, the $2J+1$ levels of the spin-orbit ground state are then split by the crystal's electric field -- an effect that is on the order of 100~meV and typically much smaller than $\lambda J$. Thus, the crystal electric field of the rare earths can often be successfully treated as a perturbation from spin-orbit coupling. This is in contrast to transition metal systems, where a perturbative approach fails because the strength of the spin-orbit coupling is comparable or smaller than the crystal electric field splitting. 

\begin{figure}[tbp]
\includegraphics[width=5in]{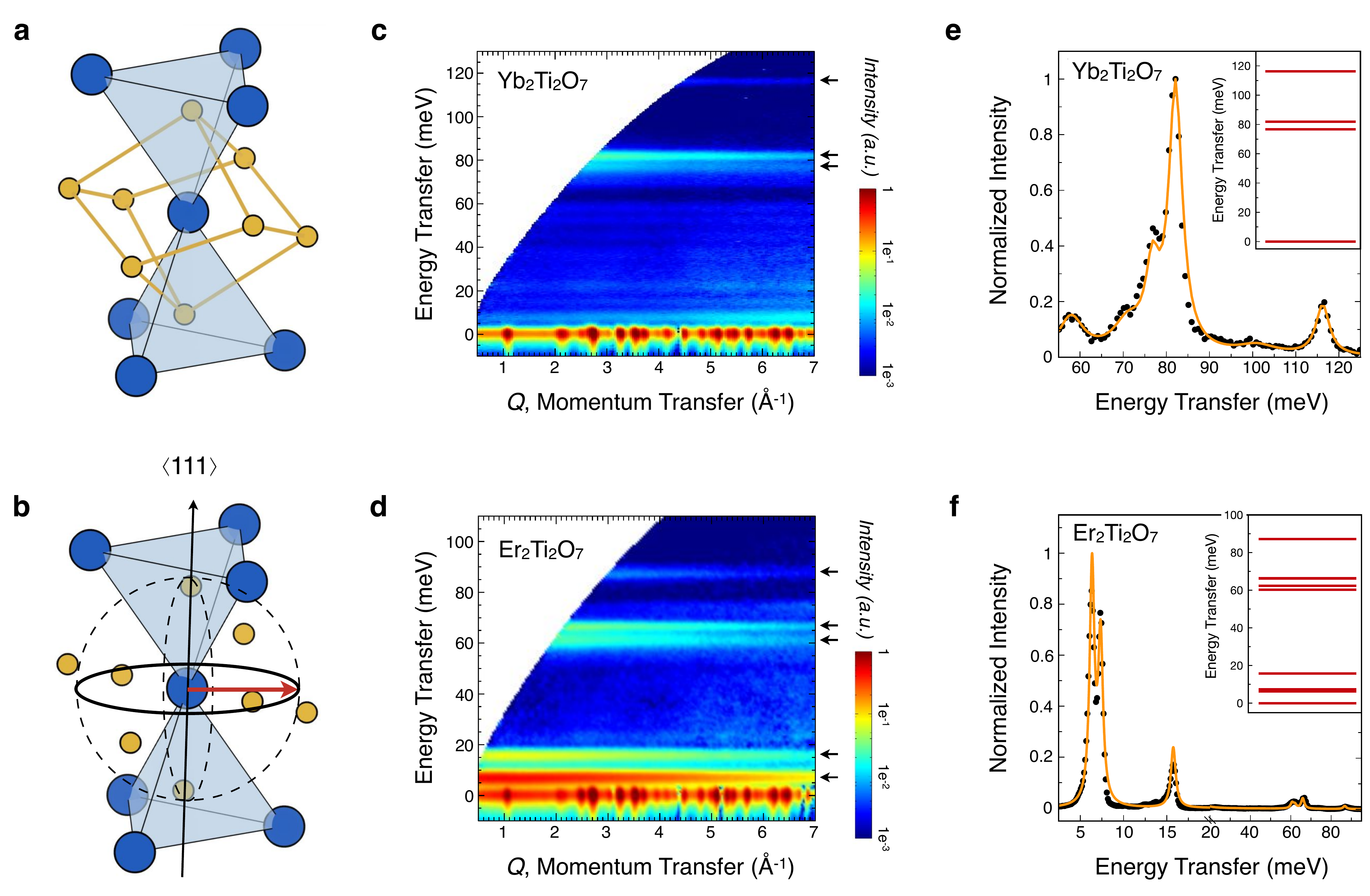}
\caption{(a) The rare earth site (blue) of the pyrochlore lattice is enclosed by eight oxygen anions (yellow), which form a distorted cube. (b) Spins with XY anisotropy are constrained to lie in the plane perpendicular to the local [111] axis. Inelastic neutron scattering measurements of (c) Yb$_2$Ti$_2$O$_7$ and (d) Er$_2$Ti$_2$O$_7$, used to determine their crystal electric field schemes. The black arrows indicate the positions of the crystal electric field transitions. The crystal electric field Hamiltonian, given in Eqn.~\ref{CEFHamiltonian}, is used to fit the experimentally measured spectra for (e) Yb$_2$Ti$_2$O$_7$~\cite{gaudet2015neutron} and (f) Er$_2$Ti$_2$O$_7$~\cite{gaudet2017effect}.}
\label{CrystalField}
\end{figure}

The crystal electric field, which lifts the $2J+1$ degeneracy of the spin-orbit ground state, depends on the point-group symmetry at the rare earth site and originates most strongly from the oxygen anions. In the pyrochlore lattice, the rare earth cation sits at the center of an eight-fold coordinate oxygen environment that forms a distorted cube, compressed along its diagonal~\cite{subramanian1983oxide}, as shown in \textbf{Figure \ref{CrystalField}(a)}. The axis along which the cube is compressed is the one that connects the oxygens at the centers of adjacent tetrahedra. This axis is the local $z$ axis and is commonly referred to as the local [111] direction. The symmetry at the rare earth site is given by the $D_{3d}$ point group, which has two-fold and three-fold rotation axes and an inversion center, defined with respect to the [111] axis. In these coordinates, a spin with Ising anisotropy will be oriented along its local [111] direction. Likewise, a spin with XY anisotropy will be constrained to lay in the plane perpendicular to [111], as shown in \textbf{Figure \ref{CrystalField}(b)}, where the local $x$ and $y$ axes are defined by the two-fold rotation axes. A peculiar aspect of the pyrochlore lattice, and one worth emphasizing, is that both of these anisotropies are explicitly defined with respect to a local direction, not a global one. Thus, Ising spins are not parallel to one another and XY spins do not share a common plane. 

\begin{marginnote}[]
\entry{Kramer's theorem}{Every energy level of a time reversal symmetric system with half integer spin must be at least doubly degenerate~\cite{kramers1930theorie}.}
\end{marginnote}

XY anisotropy is achieved on the pyrochlore lattice when the rare earth site is occupied by either ytterbium or erbium. Both of these cations, Yb$^{3+}$ and Er$^{3+}$, have an odd number of $f$ electrons, and accordingly their crystal electric field levels cannot be split beyond a doublet, as required by Kramer's theorem. Yb$^{3+}$ with $J=\frac{7}{2}$ has its lowest energy spin-orbit manifold comprised of four doublets, while Er$^{3+}$ with $J=\frac{15}{2}$ has eight doublets. Transitions between the crystal electric field levels can be directly probed by inelastic neutron scattering. Taking Yb$_2$Ti$_2$O$_7$ and Er$_2$Ti$_2$O$_7$ as examples, \textbf{Figures \ref{CrystalField}(c)} and \textbf{\ref{CrystalField}(d)} show inelastic neutron scattering maps where the crystal field transitions are indicated by the black arrows on the right hand side. As the crystal electric field is a single ion property, crystal field excitations lack dispersion and their intensities fall off with $Q$ according to the single ion magnetic form factor. While we only show crystal field schemes for Yb$_2$Ti$_2$O$_7$ and Er$_2$Ti$_2$O$_7$, the general features can be expanded to their respective families~\cite{gaudet2015neutron,hallas2016xy,gaudet2017effect}. In particular, the ytterbium pyrochlores typically have their ground states separated from their first excited crystal field state by $\sim800$~K, while the erbium pyrochlores have their ground states separated from their first excited states by $\sim80$~K.

\begin{marginnote}[]
\entry{Stevens Operators}{Formalism that uses the Wigner-Eckart theorem to express the crystal field interaction in terms of angular momentum operators \cite{stevens1952matrix}.}
\end{marginnote}

The crystal field eigenstates and eigenvalues can be determined by fitting the experimental neutron scattering spectra with the crystal electric field Hamiltonian~\cite{hutchings1964point},
\begin{equation}
\mathcal{H}_{CEF}= B^0_2\hat{O}^0_2 + B^0_4\hat{O}^0_4 + B^3_4\hat{O}^3_4 + B^0_6\hat{O}^0_6 + B^3_6\hat{O}^3_6 + B^6_6\hat{O}^6_6,
\label{CEFHamiltonian}
\end{equation}
which is based upon the point group symmetry at the rare earth site, $D_{3d}$, and is written in the Stevens operator formalism~\cite{stevens1952matrix}. 
The values of the six crystal electric field parameters ($B^0_2$, $B^0_4$, \emph{etc.}) can be systematically varied to achieve the best agreement with the experiential spectra. The result of such fittings are shown in \textbf{Figures~\ref{CrystalField}(e)} and \textbf{\ref{CrystalField}(f)} for Yb$_2$Ti$_2$O$_7$ and Er$_2$Ti$_2$O$_7$, respectively. A more detailed treatment of these types of analyses applied to pyrochlore materials can be found elsewhere~\cite{rosenkranz2000crystal,bertin2012crystal,princep2013crystal,gaudet2015neutron,ruminy2016crystal}. The result, however, is that for all the Yb and Er pyrochlores that have been measured to date, the anisotropic $g$-tensors for their ground state doublets have a larger perpendicular component, $g_{\perp}$, than parallel, $g_{\parallel}$ (\textbf{Table~\ref{tab1}}). Hence, their spin anisotropy is XY-like~\cite{hodges2001crystal,dasgupta2006crystal,cao2009ising,bertin2012crystal,guitteny2013palmer,gaudet2015neutron,hallas2016xy,gaudet2017effect}. The values of the $g$-tensors in the crystal electric field ground state are substantially reduced from the free ion value - corresponding to a reduced magnetic moment. Furthermore, the ground state eigenfunctions of the Yb and Er pyrochlores have large components of $m_J = \pm \frac{1}{2}$, which favors quantum fluctuations. 

\begin{marginnote}[]
\entry{$g$-tensor}{A generalization of the magnetic $g$-factor that, when coupled with its total angular momentum, defines the magnetic moment of an atom.}
\end{marginnote}

\begin{table}[bp]
\tabcolsep7.5pt
\caption{Ground state magnetic properties for the ytterbium and erbium pyrochlores.}
\label{tab1}
\begin{center}
\begin{tabular}{@{}l|c|c|c|c|c|c@{}}
\hline
 & $a^{\rm a}$ & $\theta_{CW}$ & $g_{\perp}/g_{\parallel}$ & $T^*$ & $T_C$ or $T_N$ & Ordered State\\
 & (\AA) & (K) & & (K) & (K) & \\
\hline
Yb$_2$Ge$_2$O$_7$ & 9.8 \cite{shannon1968synthesis} & 0.9 \cite{dun2014chemical} & 1.7 \cite{hallas2016xy} & 3.7 \cite{hallas2016universal} & 0.6 \cite{dun2015antiferromagnetic} & $\psi_2$ or $\psi_3$ AFM \cite{hallas2016xy,dun2015antiferromagnetic}\\
Yb$_2$Ti$_2$O$_7$ & 10.0 \cite{lau2006stuffed} & 0.8 \cite{hodges2002first} & 1.9 \cite{gaudet2015neutron} & 2.5 \cite{hodges2002first} & 0.25 \cite{hodges2002first} & $\Gamma_9$ FM$^{\rm b}$ \cite{yasui2003ferromagnetic,gaudet2016gapless}\\
Yb$_2$Pt$_2$O$_7$ & 10.1 \cite{sleight1968new} & 0.9 \cite{cai2016high} & -- & 2.4 \cite{cai2016high} & 0.30 \cite{cai2016high} & Undeterm. FM \cite{cai2016high}\\
Yb$_2$Sn$_2$O$_7$ & 10.3 \cite{kennedy1997structural} & 0.5 \cite{matsuhira2002low} & -- & 1.8 \cite{dun2013yb} & 0.15 \cite{yaouanc2013dynamical} & $\Gamma_9$ FM \cite{yaouanc2013dynamical,lago2014glassy}\\
\hline
Er$_2$Ge$_2$O$_7$ & 9.9 \cite{shannon1968synthesis} & $-22$ \cite{li2014long} & 3.3 \cite{gaudet2017effect} & -- & 1.4 \cite{dun2015antiferromagnetic} & $\psi_2$ or $\psi_3$ AFM$^{\rm c}$ \cite{dun2015antiferromagnetic} \\
Er$_2$Ti$_2$O$_7$ & 10.1 \cite{lau2006stuffed} & $-22$ \cite{bramwell2000bulk} & 1.6 \cite{gaudet2017effect} & -- & 1.2 \cite{de2012magnetic} & $\psi_2$ AFM \cite{poole2007magnetic}\\
Er$_2$Pt$_2$O$_7$ & 10.1 \cite{sleight1968new} & $-22$ \cite{cai2016high} & 28 \cite{gaudet2017effect} & 1.5 \cite{cai2016high} & 0.30 \cite{cai2016high} & $\Gamma_7$ PC AFM \cite{hallas2017phase}\\
Er$_2$Sn$_2$O$_7$ & 10.4 \cite{kennedy1997structural} & $-14$ \cite{matsuhira2002low} & 54 \cite{gaudet2017effect} & 4.3 \cite{ghamdi2014thermodynamic} & 0.11 \cite{petit2017long} & $\Gamma_7$ PC AFM \cite{petit2017long}\\
\hline
\end{tabular}
\end{center}
\begin{tabnote}
$^{\rm a}$Measured at room temperature and truncated at one decimal places; $^{\rm b}$The low temperature magnetism of Yb$_2$Ti$_2$O$_7$ is sample dependent, as we elaborate on in Section 4.2; $^{\rm c}$This work concluded the ground state of Er$_2$Ge$_2$O$_7$ is $\psi_3$ based on an incorrect assumption, as we elaborate in the main text.
\end{tabnote}
\end{table}

\section{GROUND STATE PHASE DIAGRAM OF XY PYROCHLORES}

Due to their crystal electric field phenomenology, the XY pyrochlore magnets Yb$_2B_2$O$_7$ and Er$_2B_2$O$_7$ have well-isolated ground state doublets. Consequently, their low temperature magnetism is well described by an effective $S=\frac{1}{2}$ system with a reduced magnetic moment and an anisotropic XY $g$-tensor. A minimal model that can describe many of their observed magnetic properties is the anisotropic nearest-neighbor exchange Hamiltonian~\cite{yan2017theory}: 
\begin{equation}
\mathcal{H}_{ex}= \sum_{\substack{\langle i,j \rangle}} J_{ij}^{\mu\nu}S_{i}^{\mu}S_{j}^{\nu}.
\label{ExchangeHamiltonian}
\end{equation} 
This Hamiltonian permits anisotropic exchange between the three spatial coordinates of each spin, $S_{j}^{\mu}$, resulting in an exchange matrix (rather than the more typical exchange constant) defined, in principle, by nine different components, $J_{ij}^{\mu\nu}$. However, the point group symmetry of the rare earth site in the pyrochlore lattice reduces the number of independent exchange parameters to just four~\cite{curnoe2007quantum}. Depending on the values of these four exchange parameters, the anisotropic exchange Hamiltonian can support the stabilization of a spin liquid ground state~\cite{canals1998pyrochlore,conlon2009spin,benton2016spin} as well as the emergence of classical~\cite{harris1997geometrical,den2000dipolar} or quantum spin ice ground states~\cite{hermele2004pyrochlore,banerjee2008unusual,savary2012coulombic,benton2012seeing,lee2012generic,savary2013spin,gingras2014quantum}.

\begin{marginnote}[]
\entry{Time-of-flight neutron scattering}{A technique where one or both of the incident and scattering neutron energies are determined by its speed.}
\end{marginnote}

The experimental determination of a real material's exchange parameters is a problem that is well-suited to the technique of inelastic neutron scattering. First, the spin wave dispersions of a single crystal can be mapped out along different crystallographic directions, a task that is increasingly straightforward given the capabilities of modern time-of-flight neutron instrumentation. Then, using linear spin wave theory derived from the Hamiltonian of \textbf{Equation~\ref{ExchangeHamiltonian}}, a fit of the experimental spectra can be performed until a set of exchange parameters and $g$-tensor components have been found that yield good agreement~\cite{ross2011quantum}. In principle, this approach can be applied within the long range ordered state of a material at low temperatures. However, if the material displays a disordered ground state in zero magnetic field or if the ordered state is complicated and not completely understood, linear spin wave theory can still successfully be applied in the field polarized state of the material, where its magnetic moments are tending to follow the magnetic field. 

\begin{figure}[tp]
\includegraphics[width=5in]{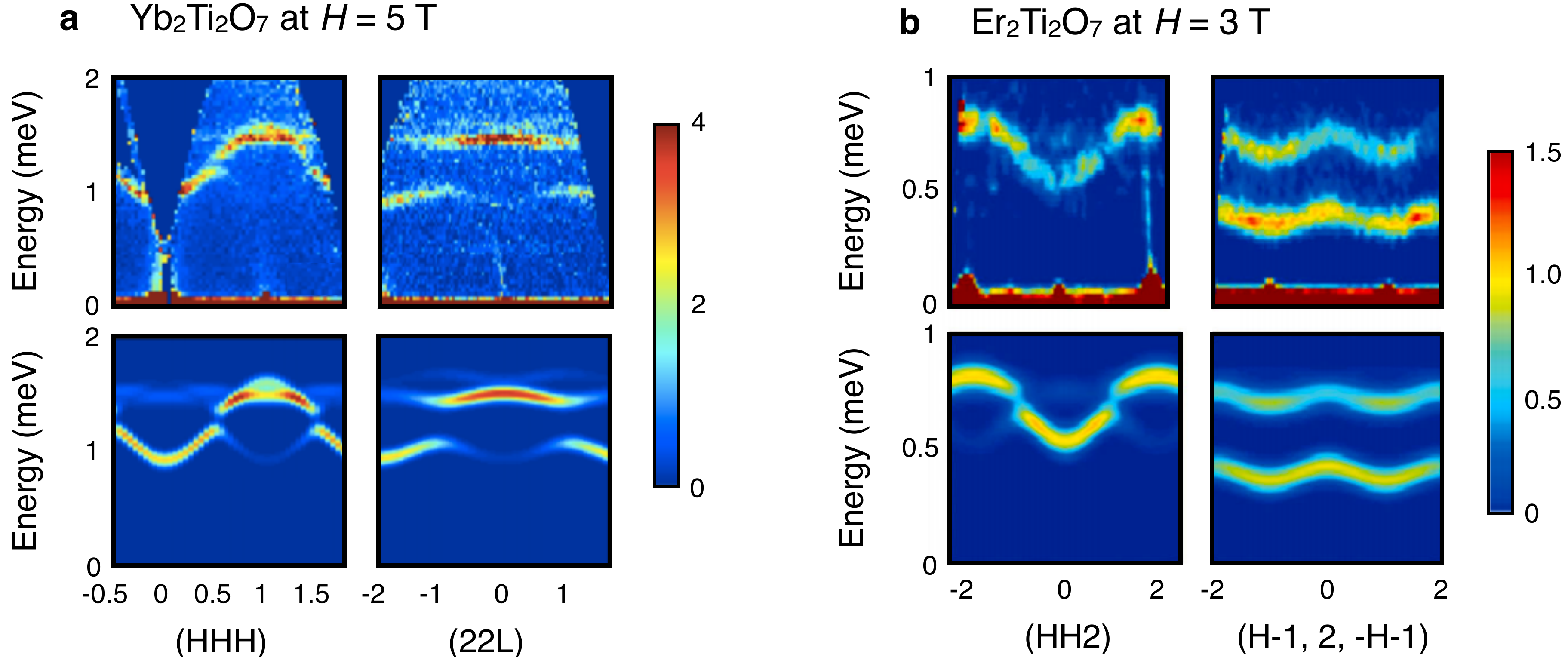}
\caption{The measured (top row) and computed (bottom row) spin wave spectra along various crystallographic directions for (a) Yb$_2$Ti$_2$O$_7$ at $H=5$~T, adapted with permission from Ref.~\cite{ross2011quantum} and (b) Er$_2$Ti$_2$O$_7$ at $H=3$~T, adapted with permission from Ref.~\cite{savary2012order}, copyrighted by the American Physical Society. The fits to the experimental spectra, which were obtained using inelastic neutron scattering, allow the determination of the anisotropic exchange couplings in the microscopic spin Hamiltonian.}
\label{SpinWaves}
\end{figure} 

\begin{figure}[tbp]
\includegraphics[width=5in]{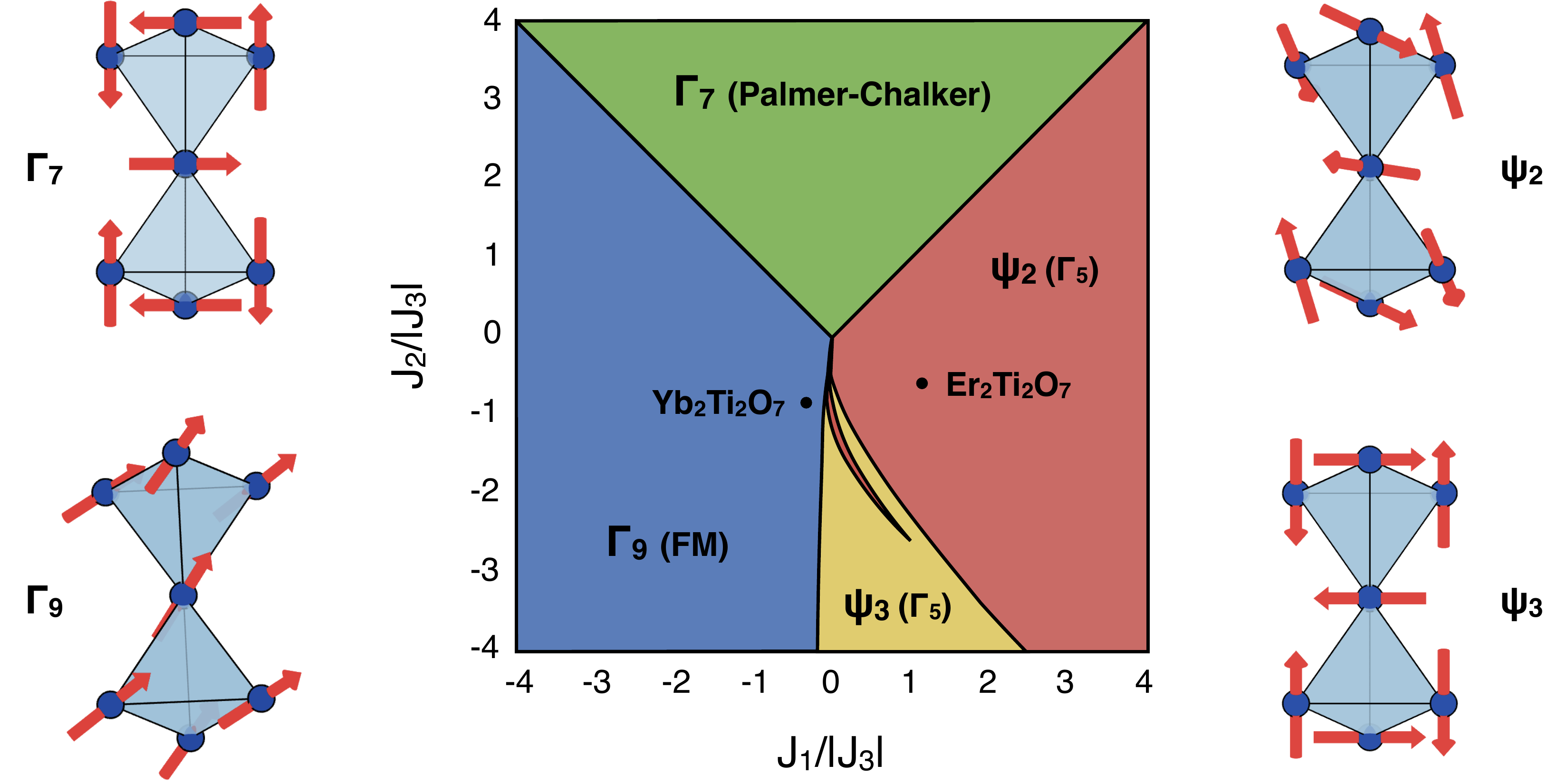}
\caption{The classical ground state phase diagram for pyrochlores with anisotropic exchange, where $J_4 = 0$ - adapted from Ref.~\cite{yan2017theory} with permission, copyrighted by the American Physical Society. This phase diagram contains four classically ordered states - the spin configurations of which are presented on two tetrahedra. The exchange parameters determined for Yb$_2$Ti$_2$O$_7$~\cite{ross2011quantum} and Er$_2$Ti$_2$O$_7$~\cite{savary2012order} place these two materials in the $\Gamma_9$ and $\psi_2$ regions, respectively.}
\label{Phasediagram}
\end{figure}

The success of this approach relies on the availability of large single crystal samples, which restricts its use to a select few materials. Amongst the family of XY pyrochlores, the titanates have provided the most successful platform for the growth of large single crystals~\cite{balakrishnan1998single,dabkowska2010crystal,li2013single}. Several groups have estimated the anisotropic exchange couplings for Yb$_2$Ti$_2$O$_7$ and Er$_2$Ti$_2$O$_7$ with inelastic neutron scattering \cite{ross2011quantum,savary2012order,petit2014order,robert2015spin,thompson2017quasiparticle}. Still, this is not a completely straightforward exercise, as there are six strongly coupled parameters to be fit, if the components of the $g$-tensor are to be adjusted in addition to the four exchange terms. \textbf{Figure~\ref{SpinWaves}} shows an example of an analysis of this type for Yb$_2$Ti$_2$O$_7$~\cite{ross2011quantum} and Er$_2$Ti$_2$O$_7$~\cite{savary2012order}. The precise values of the exchange parameters for Yb$_2$Ti$_2$O$_7$ vary amongst different groups, but there is consensus that its exchange parameters, at least at the mean field level, should lead to a $k = 0$, $\Gamma_9$ splayed ferromagnetic ordered state, as shown in \textbf{Figure~\ref{Phasediagram}}~\cite{thompson2011rods,ross2011quantum,robert2015spin,jaubert2015multiphase,yan2017theory,thompson2017quasiparticle}. Experimentally, a splayed ferromagnetic ground state has indeed been observed for some samples of Yb$_2$Ti$_2$O$_7$~\cite{yasui2003ferromagnetic,chang2012higgs,gaudet2016gapless,yaouanc2016novel,scheie2017lobed}. However, this state displays a number of peculiarities, one of which is that low levels of quenched disorder drastically impacts the formation of long-range magnetic order~\cite{yaouanc2011single,ross2012lightly,d2013unconventional,arpino2017impact}. The exchange parameters for Er$_2$Ti$_2$O$_7$ place it within the antiferromagnetic $\Gamma_5$ manifold ($\psi_2$ or $\psi_3$, shown in \textbf{Figure~\ref{Phasediagram}}) at the mean field level, but it has been shown that order-by-disorder and virtual crystal field transitions will select $\psi_2$~\cite{mcclarty2009energetic,savary2012order,zhitomirsky2012quantum,oitmaa2013phase,wong2013ground,bonville2013magnetization,petit2014order,javanparast2015order,rau2016order}. It is interesting to note that order-by-disorder can select $\psi_3$ for exchange parameters different than those of Er$_2$Ti$_2$O$_7$~\cite{wong2013ground,maryasin2014order}, but $\psi_2$ is uniformly favored by current models of virtual crystal field transitions~\cite{rau2016order}. Experimentally, differentiating between the $\psi_2$ and $\psi_3$ states is very challenging as their diffraction patterns are identical. They can, however, be distinguished via polarized neutron scattering measurements on a single crystal, and this has been used to confirm that the ordered phase of Er$_2$Ti$_2$O$_7$ is a pure $\psi_2$ state~\cite{poole2007magnetic}. 

Taking the experimentally determined exchange parameters of Yb$_2$Ti$_2$O$_7$ and Er$_2$Ti$_2$O$_7$ as a starting point, theoretical studies have explored the ordered phases that are expected classically within a subspace of exchange parameters relevant for the XY pyrochlores~\cite{wong2013ground,jaubert2015multiphase,yan2017theory}. The resulting phase diagram is rich, revealing the possibility of stabilizing various magnetic structures, as shown in \textbf{Figure~\ref{Phasediagram}}~\cite{yan2017theory}. This phase diagram contains four $k=0$ magnetic structures, where the spin configuration of each is shown around the perimeter. In addition to the ferromagnetic $\Gamma_9$ phase found for parameters appropriate to Yb$_2$Ti$_2$O$_7$ and the antiferromagnetic $\psi_2$ phase for Er$_2$Ti$_2$O$_7$, there are two additional antiferromagnetic states: $\psi_3$, which is part of the $\Gamma_5$ manifold, and $\Gamma_7$, the so-called Palmer-Chalker state~\cite{palmer2000order}. All these magnetic structures can be obtained within a narrow range of exchange parameters, suggesting that phase competition could exist within the family of XY pyrochlores~\cite{yan2017theory,jaubert2015multiphase}. This scenario would imply that a change of magnetic ground state amongst the Yb and Er pyrochlores would require only modest changes to the anisotropic exchange parameters, as might be induced by chemical or external pressure.

\begin{marginnote}[]
\entry{Palmer-Chalker State}{Named after the authors of Ref.~\cite{palmer2000order}, which showed that $\Gamma_7$ is the ground state for a Heisenberg pyrochlore antiferromagnet with dipolar interactions.}
\end{marginnote}

Chemical pressure can be realized by substituting a non-magnetic constituent for an element of varying size. In the case of the XY titanate pyrochlores, this leads to the replacement of titanium on the $B$ site by germanium, tin, or platinum (\textbf{Figure~\ref{PeriodicTable}(b)}). As reported in \textbf{Table~\ref{tab1}}, the magnetic structures of most of these chemical pressure analogs have already been determined. For the ytterbium pyrochlores, no change of magnetic ground state is obtained by incorporation of tin~\cite{yaouanc2013dynamical,lago2014glassy}, but an antiferromagnetic $\Gamma_5$ phase is obtained for Yb$_2$Ge$_2$O$_7$~\cite{dun2015antiferromagnetic,hallas2016xy}. For the erbium pyrochlores, germanium substitution also results in a $\Gamma_5$ ($\psi_2$ or $\psi_3$) ground state~\cite{dun2015antiferromagnetic}. The effect of a magnetic field on the powder diffraction pattern of Er$_2$Ge$_2$O$_7$ has been argued to indicate a pure $\psi_3$ state~\cite{dun2015antiferromagnetic}. However, the domain selection arguments upon which this determination is based have been found to be incorrect~\cite{maryasin2016low,gaudet2016magnetic,gaudet2017experimental}. 
Thus, while it is possible that Er$_2$Ge$_2$O$_7$ has a $\psi_3$ ground state, it remains to be definitively shown. For Er$_2$Pt$_2$O$_7$ and Er$_2$Sn$_2$O$_7$, powder neutron diffraction studies have uncovered $\Gamma_7$ Palmer-Chalker ground states but with significantly reduced $T_N$'s from the other erbium pyrochlores~\cite{hallas2017phase,petit2017long}. This suppression of magnetic order could originate from phase competition between the $\Gamma_5$ and $\Gamma_7$ states~\cite{guitteny2013palmer,yan2017theory,hallas2017phase}. Moreover, the ensemble of states observed within these two XY pyrochlore families confirms the proximity of the $\Gamma_5$, $\Gamma_7$, and $\Gamma_9$ states in their phase space and empirically validates the scenario of strong phase competition. This stands in contrast to the classical spin ice pyrochlores, Ho$_2$Ti$_2$O$_7$ and Dy$_2$Ti$_2$O$_7$, where germanium and tin substitution at the $B$ site leaves the ground state properties largely unchanged~\cite{matsuhira2000low,matsuhira2002low,zhou2012chemical,hallas2012statics}.

\section{EXPERIMENTAL SIGNATURES OF PHASE COMPETITION IN THE XY PYROCHLORES} 

The variety of ground states observed amongst the XY pyrochlores, combined with a theoretically-derived phase diagram based on anisotropic exchange, leads one to suspect that phase competition is likely important in these materials. In this section, we describe several experimental observations that could be manifestations of this phase competition. We first demonstrate that the ytterbium based XY pyrochlores possess unconventional spin dynamics that are clearly linked with a broad feature in their specific heat, rather than the sharp anomalies they display at $T_C$ or $T_N$. We argue that the origin of these unconventional dynamics is the proximity of nearby ordered phases and the competition between them. We further demonstrate the sensitivity of the ground state in Yb$_2$Ti$_2$O$_7$ to both weak quenched disorder and hydrostatic pressure, and argue that this too has a natural explanation in phase competition. 

\subsection{UNCONVENTIONAL SPIN DYNAMICS}

\begin{figure}[tbp]
\includegraphics[width=5in]{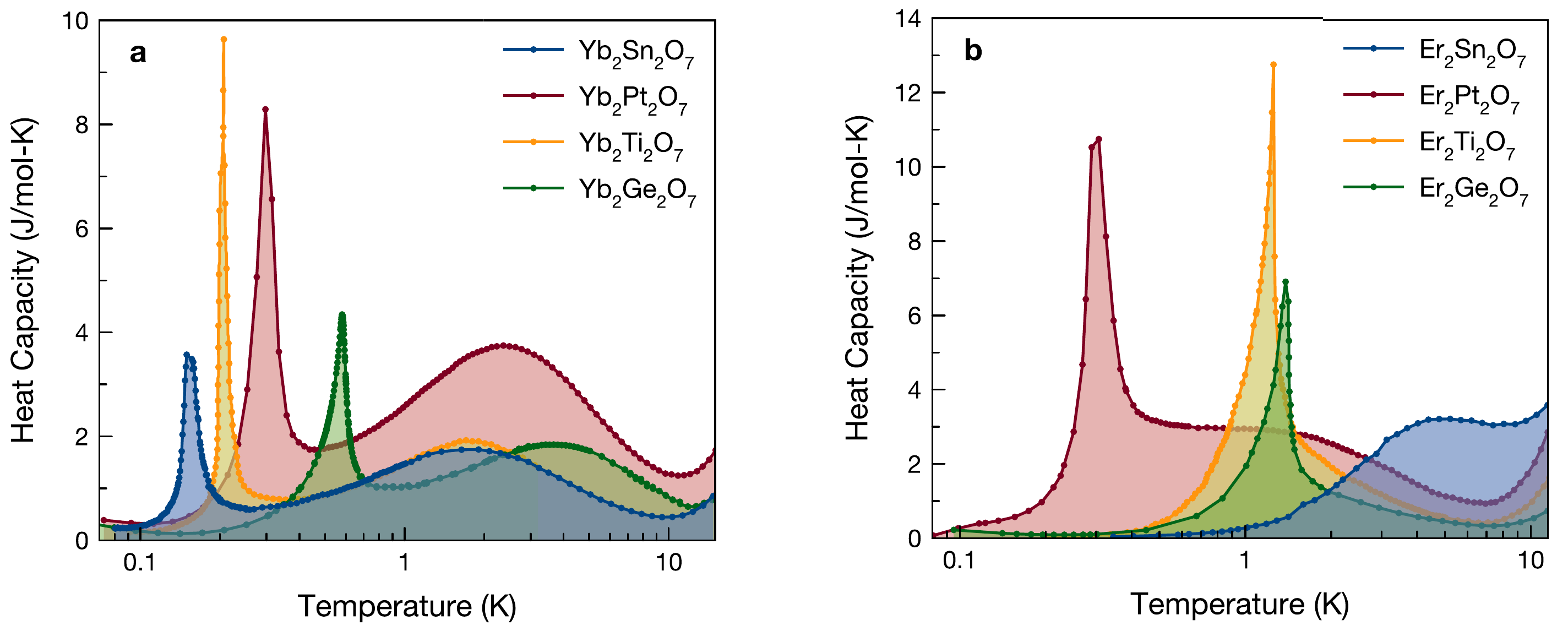}
\caption{The low temperature heat capacity of the (a) Yb$_2B_2$O$_7$ pyrochlores ($B$ = Ge \cite{dun2015antiferromagnetic}, Ti \cite{hodges2002first}, Pt \cite{cai2016high}, Sn \cite{yaouanc2013dynamical,dun2013yb}) and the (b) Er$_2B_2$O$_7$ pyrochlores ($B$ = Ge \cite{dun2015antiferromagnetic}, Ti \cite{de2012magnetic}, Pt \cite{cai2016high}, Sn \cite{ghamdi2014thermodynamic}).}
\label{HeatCapacity}
\end{figure}

A summary of heat capacity measurements performed on the ytterbium and erbium pyrochlores is shown in \textbf{Figure~\ref{HeatCapacity}}. Over the temperature range displayed, the lattice contribution is negligible, and the heat capacity is dominated by the magnetic component. There is a striking similarity in the form of the specific heat for all the ytterbium pyrochlores, as well as Er$_2$Pt$_2$O$_7$, which consists of a sharp lambda-like anomaly at $T_C$ or $T_N$ and a broad, higher temperature anomaly at a temperature we label as $T^*$~\cite{hodges2002first,yaouanc2013dynamical,dun2013yb,dun2015antiferromagnetic,cai2016high}. Two compounds order above 1~K, Er$_2$Ti$_2$O$_7$ and Er$_2$Ge$_2$O$_7$, and interestingly, these are also the only two that do not display broad specific heat anomalies~\cite{de2012magnetic,dun2015antiferromagnetic}. Lastly, Er$_2$Sn$_2$O$_7$ displays only a broad specific heat anomaly~\cite{ghamdi2014thermodynamic}, as there is no reported data that extends below its $T_N=0.11$~K ordering transition~\cite{petit2017long}, where we assume a sharp peak exists. Our conjecture is that the presence of a broad specific heat anomaly, coincident with a suppression of magnetic ordering to lower temperature, is an experimental signature of phase competition in the XY pyrochlores. It is worth noting that these broad specific heat anomalies do not have the characteristic form of a Schottky anomaly, as would be expected if their origin was the thermal depopulation of a crystal field level. As discussed in Section~2, the crystal electric field splittings in the Yb and Er pyrochlores are on the order of $\sim 800$~K and $\sim 80$~K, respectively, far above the energy scale for these broad specific heat anomalies. Furthermore, the combined entropy release of the sharp and broad anomalies is $R\cdot\ln{(2)}$~\cite{hallas2016universal,cai2016high}. Thus, the entropy release associated with these broad specific heat anomalies is related to spin degrees of freedom within the isolated ground state doublet.

The fact that Er$_2$Ti$_2$O$_7$ displays no such broad anomaly and orders at relatively high temperature indicates that phase competition is less relevant to its ground state selection. This is consistent with the placement of Er$_2$Ti$_2$O$_7$ in the classical anisotropic exchange phase diagram, relatively far from the phase boundaries of the $\Gamma_9$ and $\Gamma_7$ ordered phases (\textbf{Figure~\ref{Phasediagram}}). In contrast, each of the ytterbium pyrochlores exhibits a broad specific heat anomaly, which we propose is connected with phase competition. For the ytterbium pyrochlores, we expect the most intense phase competition to arise between the ferromagnetic $\Gamma_9$ state and the antiferromagnetic $\Gamma_5$ state -- a scenario that is consistent with the placement of Yb$_2$Ti$_2$O$_7$ in the XY phase diagram (\textbf{Figure~\ref{Phasediagram}}) and also with the fact that the ground states of the ytterbium pyrochlores are known to span these two states (\textbf{Table~\ref{tab1}}). Lastly, we suggest that the broad specific heat anomalies observed in Er$_2$Sn$_2$O$_7$ and Er$_2$Pt$_2$O$_7$ also originate from phase competition, but between the $\Gamma_5$ and the $\Gamma_7$ ordered phases. These phases and the boundary between them correspond to net antiferromagnetic interactions, consistent with the Curie-Weiss temperatures in the Er pyrochlores (\textbf{Table~\ref{tab1}}). It is also empirically supported by the fact that both Er$_2$Ti$_2$O$_7$ and Er$_2$Ge$_2$O$_7$ order into the $\Gamma_5$ phase, while Er$_2$Pt$_2$O$_7$ orders into the $\Gamma_7$ phase. 

\begin{figure}[tbp]
\includegraphics[width=5in]{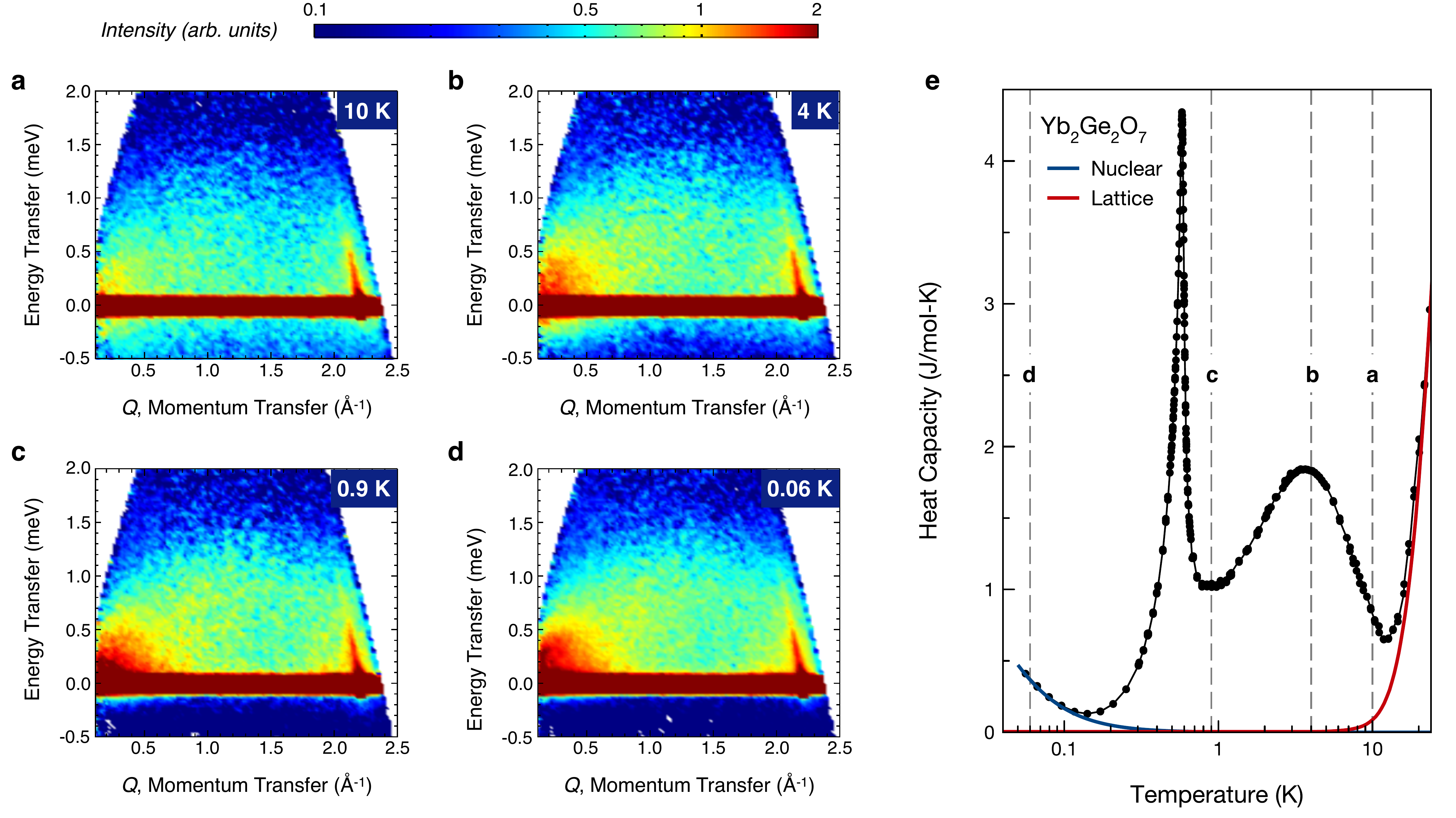}
\caption{(a-d) The evolution of the spin dynamics in Yb$_2$Ge$_2$O$_7$ as a function of temperature measured by inelastic neutron scattering, adapted with permission from Ref.~\cite{hallas2016universal}. (e) The temperatures at which the inelastic neutron scattering spectra were measured are indicated in reference to the specific heat. The broad band of ferromagnetic-like excitations form well above $T_N$, in the temperature range of the broad specific heat anomaly, $T^*$.}
\label{Temp_Dep}
\end{figure}

To better understand what is physically transpiring in the XY pyrochlores at the temperature of the broad specific heat anomaly, $T^*$, we consider evidence from inelastic neutron scattering. We take Yb$_2$Ge$_2$O$_7$ as our representative example because it has the greatest degree of temperature separation between its broad and sharp heat capacity anomalies. In \textbf{Figure~\ref{Temp_Dep}}, we present the temperature dependence of the low energy inelastic neutron scattering spectra of Yb$_2$Ge$_2$O$_7$ and compare it with the heat capacity. In contrast to the inelastic neutron scattering measurements shown in Section~2, these measurements probe far smaller energy transfers ($\sim1$~meV) at lower temperatures. Thus, they are sensitive to the collective spin dynamics, as opposed to the single ion excitations. Between 10~K and 4~K, there is a significant buildup of correlations centered towards $Q=0$, a ferromagnetic zone center. This broad band of excitations decays in intensity as it extends outwards in $Q$, but without apparent coherent propagation. These spin excitations form at $\sim4$~K, well above $T_N$, and are thus the origin of the broad specific heat anomaly. The confluence of $T^*$ with a change in form of the spin dynamics is a shared feature amongst all the Yb-based pyrochlores that have been measured thus far, and also Er$_2$Pt$_2$O$_7$ and Er$_2$Sn$_2$O$_7$~\cite{sarte2011absence,dun2013yb,gaudet2016gapless,hallas2016universal,hallas2017phase,petit2017long}. That is to say, in all the XY pyrochlores that display both broad and sharp specific heat anomalies, the excitations develop some coherence at $T^*$, a temperature well-above $T_C$ or $T_N$. 

\begin{figure}[tbp]
\includegraphics[width=5in]{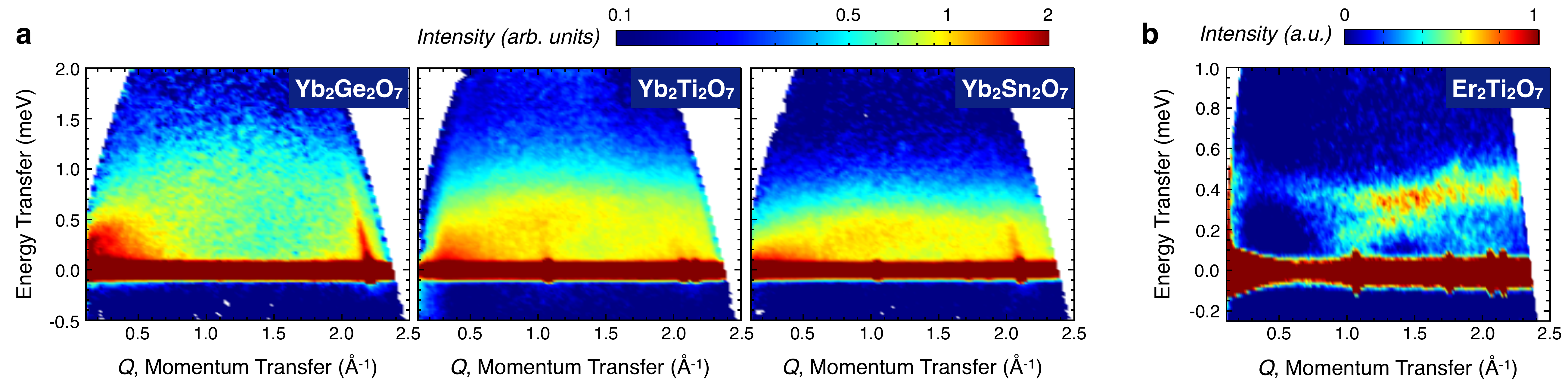}
\caption{Comparison of the low temperature (below $T_C$ or $T_N$) spin dynamics among the (a) ytterbium pyrochlores and (b) Er$_2$Ti$_2$O$_7$ as measured by inelastic neutron scattering. Experimental details for each of these data sets can be found in the corresponding original publication: (a) Yb$_2$Ge$_2$O$_7$~\cite{hallas2016universal}, (b) Yb$_2$Ti$_2$O$_7$~\cite{gaudet2016gapless}, (c) Yb$_2$Sn$_2$O$_7$~\cite{dun2013yb}, and (d) Er$_2$Ti$_2$O$_7$~\cite{ruff2008spin}. Note that Yb$_2$Ge$_2$O$_7$ and Er$_2$Ti$_2$O$_7$ share the same antiferromagnetic ordered state.}
\label{LowEnergyInelastic}
\end{figure}

As Yb$_2$Ge$_2$O$_7$ is further cooled to 0.9~K and 0.06~K, below $T_N$, there is little significant redistribution of the spectral weight. Despite ordering into the $\Gamma_5$ antiferromagnetic representation manifold, there is no corresponding response in the spin dynamics (within an energy resolution of $\sim$0.09~meV). Once again, this insensitivity of the spin dynamics to long range magnetic order is shared amongst all the ytterbium pyrochlores~\cite{dun2013yb,gaudet2016gapless,hallas2016universal}. In \textbf{Figure~\ref{LowEnergyInelastic}} we present the low temperature inelastic neutron scattering spectra for the three ytterbium pyrochlores and Er$_2$Ti$_2$O$_7$. Of all the neutron inelastic spectra shown in \textbf{Figure~\ref{LowEnergyInelastic}}, Er$_2$Ti$_2$O$_7$ is the only XY pyrochlore where the formation of the spin wave excitations coincides with the observation of long range magnetic order~\cite{ruff2008spin}. A striking difference is observed between the low temperature inelastic spectra of Yb$_2$Ge$_2$O$_7$ and Er$_2$Ti$_2$O$_7$, despite the fact that both order into a $\Gamma_5$ manifold. Instead of spin waves originating from the antiferromagnetic zone center, as observed in Er$_2$Ti$_2$O$_7$, Yb$_2$Ge$_2$O$_7$ has its largest spectral weight near $Q=0$, a ferromagnetic zone center~\cite{hallas2016universal}. 

Finally, comparing the low temperature inelastic neutron spectra of the three ytterbium pyrochlores in \textbf{Figure~\ref{LowEnergyInelastic}}, it is clear that they all share a common form for their spin excitation spectra. In each Yb pyrochlore, the zero field spin excitations are very unconventional, displaying a broad continuum of gapless excitations to temperatures as low as 60~mK. The only substantial difference is the energy band-width of the spin excitations, which reduces as expected with increasing lattice parameter~\cite{hallas2016universal}. The anisotropic exchange parameters for Yb$_2$Ti$_2$O$_7$ should result in a $\Gamma_9$ ferromagnetic ordered state at the mean field level~\cite{ross2011quantum,robert2015spin,thompson2017quasiparticle}, and this expectation has been partially fulfilled in some samples of Yb$_2$Ti$_2$O$_7$~\cite{yasui2003ferromagnetic,chang2012higgs,gaudet2016gapless,yaouanc2016novel,scheie2017lobed}. Thus, at zero magnetic field, well defined spin waves and a relatively large spin wave gap are predicted to emerge at $T_C$~\cite{gaudet2016gapless,yan2017theory,thompson2017quasiparticle}. However, the zero field spin excitations reflect a far more disordered state. The evolution of this continuum of scattering at zero field into dispersing magnons at high field has recently been investigated in Yb$_2$Ti$_2$O$_7$~\cite{thompson2017quasiparticle}. In magnetic fields up to 9~T, the intensity from two-magnon processes is strongly underestimated by a non-interacting spin-wave model, suggesting the presence of strong quantum fluctuations that persist even in high magnetic fields. At lower fields, there is an overlap between one and two-magnon states leading to strong renormalization effects and quasi-particle breakdown in zero field~\cite{pan2014low,pan2016measure,thompson2017quasiparticle}. Thus, the measured spin excitation spectrum of Yb$_2$Ti$_2$O$_7$ bears little resemblance to that predicted within the low temperature mean field ordered state.

\subsection{SENSITIVITY TO DISORDER}

\begin{figure}[tbp]
\includegraphics[width=5in]{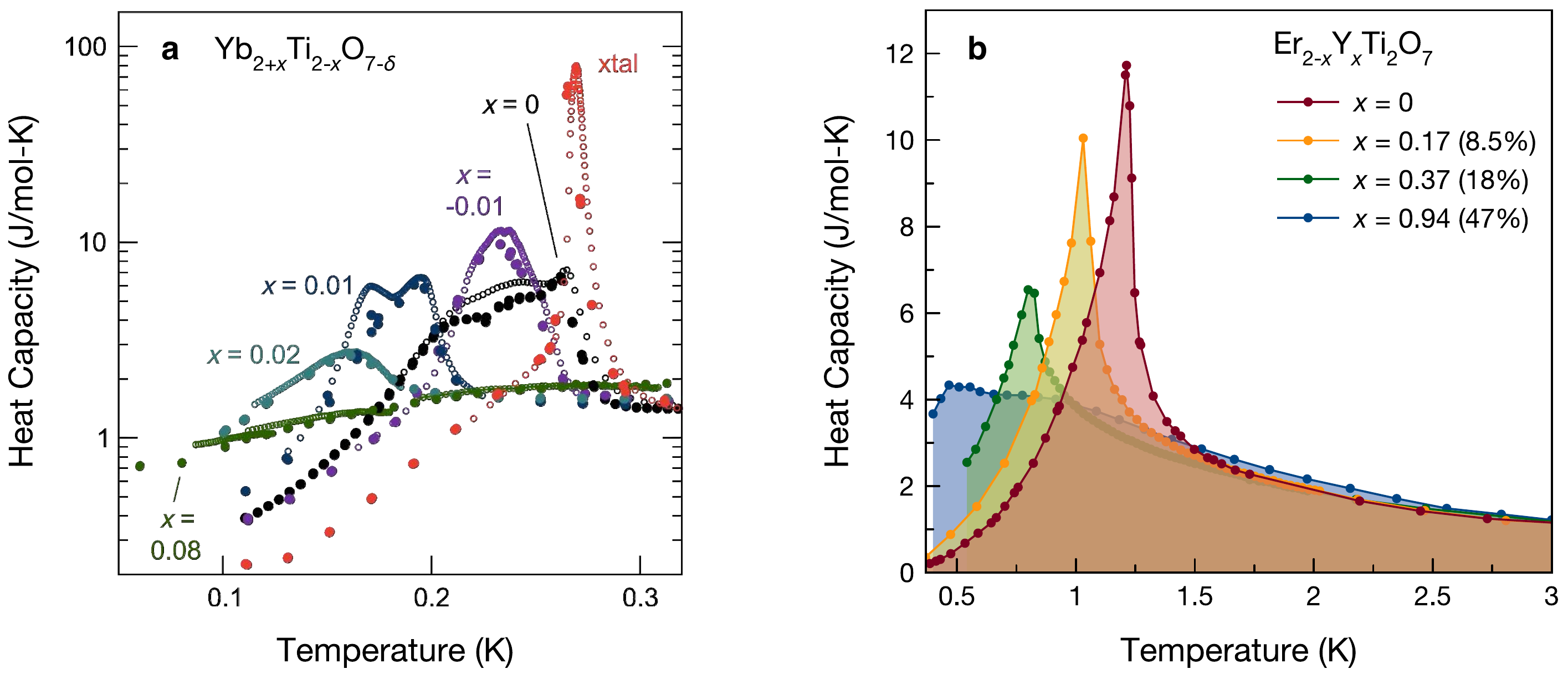}
\caption{The effect of quenched disorder on the magnetic properties of (a) Yb$_2$Ti$_2$O$_7$, adapted with permission from Ref.~\cite{arpino2017impact}, copyrighted by the American Physical Society and (b) Er$_2$Ti$_2$O$_7$, adapted with permission from Ref.~\cite{niven2014magnetic}. Yb$_2$Ti$_2$O$_7$ exhibits a pronounced sensitivity to small levels of off-stoichiometry while the long-range order of Er$_2$Ti$_2$O$_7$ is robust to disorder.}
\label{SampleDependence}
\end{figure}

Within the family of XY pyrochlores, only the titanates, Yb$_2$Ti$_2$O$_7$ and Er$_2$Ti$_2$O$_7$, have long existed as large single crystals. Thus, for these two materials, comparisons can be made between polycrystalline materials, prepared by solid state synthesis, and single crystals, grown by the floating zone technique. This comparison has revealed that the magnetic ground state of Yb$_2$Ti$_2$O$_7$ has a fascinating dependence on quenched disorder. This unusual sample dependence was first reported in early heat capacity studies~\cite{yaouanc2011single,ross2011dimensional}, and can be broadly summarized as polycrystalline samples possessing sharper heat capacity anomalies at relatively high $T_C$'s, while single crystal samples possess much broader anomalies, sometimes multiple peaks, and with considerably lower $T_C$'s~\cite{yaouanc2011single,ross2011dimensional,d2013unconventional,chang2014static,lhotel2014first}. Subsequent characterization with neutron diffraction has revealed the likely origin of this effect: floating zone single crystals grown from stoichiometric precursors have a small concentration of anti-site defects. Specifically, a small fraction of Yb$^{3+}$ cations, on the order of 1\%, can be stuffed onto the $B$ site, leading to a net off-stoichiometry, Yb$_{2+x}$Ti$_{2-x}$O$_{7-\delta}$~\cite{ross2012lightly}, which is necessarily accompanied by oxygen defects~\cite{sala2014vacancy,mostaed2017atomic}. Very recently, a systematic investigation of off-stoichiometry in Yb$_2$Ti$_2$O$_7$ has been carried out, as shown in \textbf{Figure~\ref{SampleDependence}(a)}~\cite{arpino2017impact}. It is clear that small levels of quenched disorder give a large and pronounced suppression of $T_C$, consistent with earlier studies, but now with the off-stoichiometry being introduced in a controlled manner. This same work also employed the traveling solvent floating zone method to successfully grow a stoichiometric single crystal of Yb$_2$Ti$_2$O$_7$ that exhibits a sharp heat capacity anomaly and a high $T_C$, also shown in \textbf{Figure~\ref{SampleDependence}(a)}~\cite{arpino2017impact}. This sample variability, which is most conveniently demonstrated by heat capacity, has also been appreciated with muon spin relaxation ($\mu$SR) and neutron scattering, where some samples display the signatures of long range magnetic order~\cite{yasui2003ferromagnetic,chang2012higgs,chang2014static,yaouanc2016novel,gaudet2016gapless,scheie2017lobed} and others do not~\cite{hodges2002first,gardner2004spin,ross2009two,d2013unconventional,robert2015spin,kermarrec2017ground}.

\begin{marginnote}[]
\entry{Floating Zone Crystal Growth}{A crucible-free, melt based technique that can produce large, high quality single crystals.}
\end{marginnote}

The sensitivity to quenched disorder observed in Yb$_2$Ti$_2$O$_7$ is {\it not} seen in the low temperature phase behavior of Er$_2$Ti$_2$O$_7$~\cite{de2012magnetic}. This is particularly interesting given that one would expect roughly the same type and concentration of disorder to be present in samples of each. To our knowledge, no sample dependence or variation between polycrystalline and single crystal samples of Er$_2$Ti$_2$O$_7$ has been reported. In fact, systematic studies of magnetic dilution in Er$_2$Ti$_2$O$_7$, where Er$^{3+}$ is substituted with non-magnetic Y$^{3+}$, show that dilution gradually suppresses $T_N$, consistent with the expectations of three dimensional percolation theory (\textbf{Figure~\ref{SampleDependence}(b)})~\cite{niven2014magnetic}. However, while it has been shown that the $\Gamma_5$ state is robust to disorder~\cite{gaudet2016magnetic}, there are theoretical proposals that magnetic dilution may induce a transition from $\psi_2$ to $\psi_3$, without additional suppression of $T_N$~\cite{maryasin2014order,andreanov2015order}. Regardless, the insensitivity of the $\Gamma_5$ magnetically ordered state in Er$_2$Ti$_2$O$_7$ to impurities is consistent with the idea that it is far-removed from competing XY phases~\cite{yan2017theory}. Likewise, the astounding sensitivity of the ground state magnetism of Yb$_2$Ti$_2$O$_7$ to low levels of disorder or off-stoichiometry is qualitatively understood by its proximity to competing XY phases~\cite{jaubert2015multiphase,yan2017theory}. Indeed, small changes in stoichiometry could slightly vary the anisotropic exchange terms, which would naturally be expected to have a larger impact in the presence of proximate phases.

\subsection{SENSITIVITY TO APPLIED HYDROSTATIC PRESSURE}

\begin{figure}[tbp]
\includegraphics[width=5in]{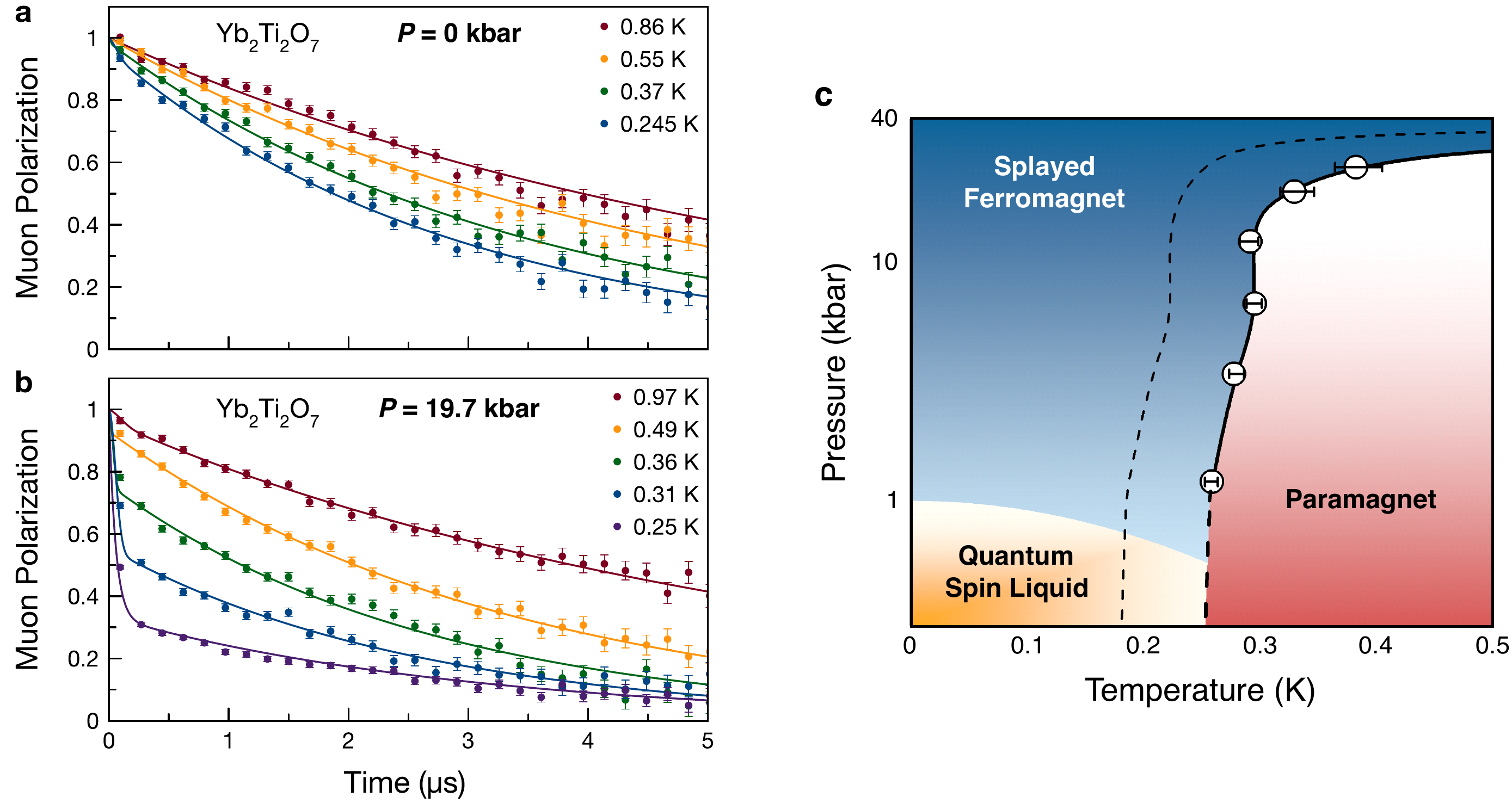}
\caption{Muon spin relaxation measurements showing that a sample of Yb$_2$Ti$_2$O$_7$ that does not exhibit long range magnetic order at (a) ambient pressure can have magnetic order induced by the application of a modest (b) applied pressure of 19.7~kbar. (c) The resulting phase diagram for this sample of Yb$_2$Ti$_2$O$_7$ showing how pressure can be used as a tuning parameter from splayed ferromagnet to quantum spin liquid. Adapted from Ref.~\cite{kermarrec2017ground} with permission.}
\label{Pressure_muSR}
\end{figure}

A corollary to the sensitivity of Yb$_2$Ti$_2$O$_7$'s ground state to quenched disorder, which can be thought of as inducing a chemical pressure, is that applied pressure may also be expected to have a disproportionate effect on ground state selection. The only other pyrochlore titanate with pronounced sensitivity to low levels of disorder and off-stoichiometry, Tb$_2$Ti$_2$O$_7$~\cite{taniguchi2013long,kermarrec2015gapped}, shows long range order induced out of its enigmatic spin liquid state by the application of external pressure~\cite{mirebeau2002pressure,mirebeau2004pressure}. Thus, a natural question that arises is, what effect would externally applied pressure have on a sample of Yb$_2$Ti$_2$O$_7$ that does not exhibit the signatures of static magnetic order? The answer to this question is given in \textbf{Figure~\ref{Pressure_muSR}(a)} and \textbf{(b)}, which presents the zero field muon decay asymmetry as a function of temperature for a stoichiometric powder sample of Yb$_2$Ti$_2$O$_7$ in ambient pressure and at $P=19.7$~kbar~\cite{kermarrec2017ground}. With zero applied hydrostatic pressure, the form of the $\mu$SR time dependence displays little variation with temperature, maintaining a weak exponential fall-off for all temperatures, above and below $T_C$, indicating a magnetically disordered ground state. However, with a hydrostatic pressure of 19.7 kbar, the $\mu$SR signal changes dramatically passing through $T_C = 0.265$~K. There is a rapid relaxation of the muon polarization at early times, indicative of static internal fields and a magnetically {\it ordered} state. Complementary neutron diffraction measurements revealed that at $P=12$~kbar this same sample of Yb$_2$Ti$_2$O$_7$ undergoes a phase transition to a weakly-splayed ferromagnetic state~\cite{kermarrec2017ground}. 

\begin{marginnote}[]
\entry{Muon spin relaxation}{A technique that uses spin polarized muons as a local probe of the distribution and dynamics of internal magnetic fields in materials.}
\end{marginnote}

These measurements on Yb$_2$Ti$_2$O$_7$ are summarized in a temperature-pressure phase diagram, \textbf{Figure~\ref{Pressure_muSR}(c)}~\cite{kermarrec2017ground}. For this sample, even the smallest applied pressure, $P=1.4$~kbar, results in a long range ordered state. However, why the ground state selection of Yb$_2$Ti$_2$O$_7$ should be so sensitive to applied pressure remains a mystery, likely interconnected to the mystery that surrounds why small levels of stuffing and other defects are so effective at disrupting what would otherwise be a simple ordered state: a splayed ferromagnet. The leading contenders to resolve these mysteries are that pressure, either applied or chemically-induced by impurities, shifts terms in the anisotropic exchange Hamiltonian of Yb$_2$Ti$_2$O$_7$, pushing the system from one phase to a nearby competing ordered structure within its classical phase diagram, as illustrated in \textbf{Figure~\ref{Phasediagram}}~\cite{yan2017theory,jaubert2015multiphase}. Another factor is that the low level of Yb$^{3+}$ ions that sit at defect sites in Yb$_2$Ti$_2$O$_7$ will still carry magnetic moments, but due to the changed oxygen environment, will have Ising rather than XY anisotropy~\cite{gaudet2015neutron}. Such defect moments may well be surprisingly effective in frustrating even simple ferromagnetic order. Of course, both of these effects could be relevant, possibly in combination with other factors.

\section{XY PYROCHLORES BEYOND THE RARE EARTHS}

The availability of large single crystals allows significantly more sophisticated measurements and analyses to be performed, which generally advances the physical understanding of a given material. As has been established in the prior sections, the ability to grow large single crystals of Yb$_2$Ti$_2$O$_7$ and Er$_2$Ti$_2$O$_7$ has allowed the determination of their microscopic spin Hamiltonians. Such studies have thus far not been possible for the germanium, platinum, and tin based pyrochlores. The synthesis barrier to overcome in the case of the $A_2$Ge$_2$O$_7$ and $A_2$Pt$_2$O$_7$ pyrochlores is quite high, as these materials are prepared under high pressure conditions (of order GPa)~\cite{shannon1968synthesis,sleight1968new}. In the case of the $A_2$Sn$_2$O$_7$ pyrochlores, the dilemma is the volatility of SnO$_2$ at high temperatures, which renders the melt-based techniques utilized in the titanate pyrochlores unsuitable. However, progress has recently been made with flux growth techniques - resulting in millimeter sized crystals of $A_2$Sn$_2$O$_7$ pyrochlores~\cite{prabhakaran2016crystal}. 

Beyond the rare earth pyrochlores, and indeed beyond oxides, recent progress has been made in the growth of fluoride pyrochlores, $AA'B_2$F$_7$. In these fluoride pyrochlores, the magnetic ions are $3d$ transition metals that sit on the $B$ site and the nonmagnetic $A/A'$ site has mixed occupancy to preserve charge neutrality. A handful of these fluoride materials have recently been synthesized, such as NaCaNi$_2$F$_7$, NaSrMn$_2$F$_7$, and NaCaFe$_2$F$_7$, and there is great promise for expansion of this family \cite{krizan2014nacaco,krizan2015nacan,krizan2015nasrco2f7,sanders2016nasrmn2f7}. These materials present an exciting opportunity as they incorporate magnetic cations that are not generally accessible with oxide pyrochlores. Furthermore, the magnetic ions are transition metals, rather than rare earths, and hence the exchange interactions are stronger and the characteristic temperature scales are higher. Moreover, these new fluoride pyrochlores are grown using the floating zone method and are thus readily available as large single crystals. 

\begin{figure}[tbp]
\includegraphics[width=5in]{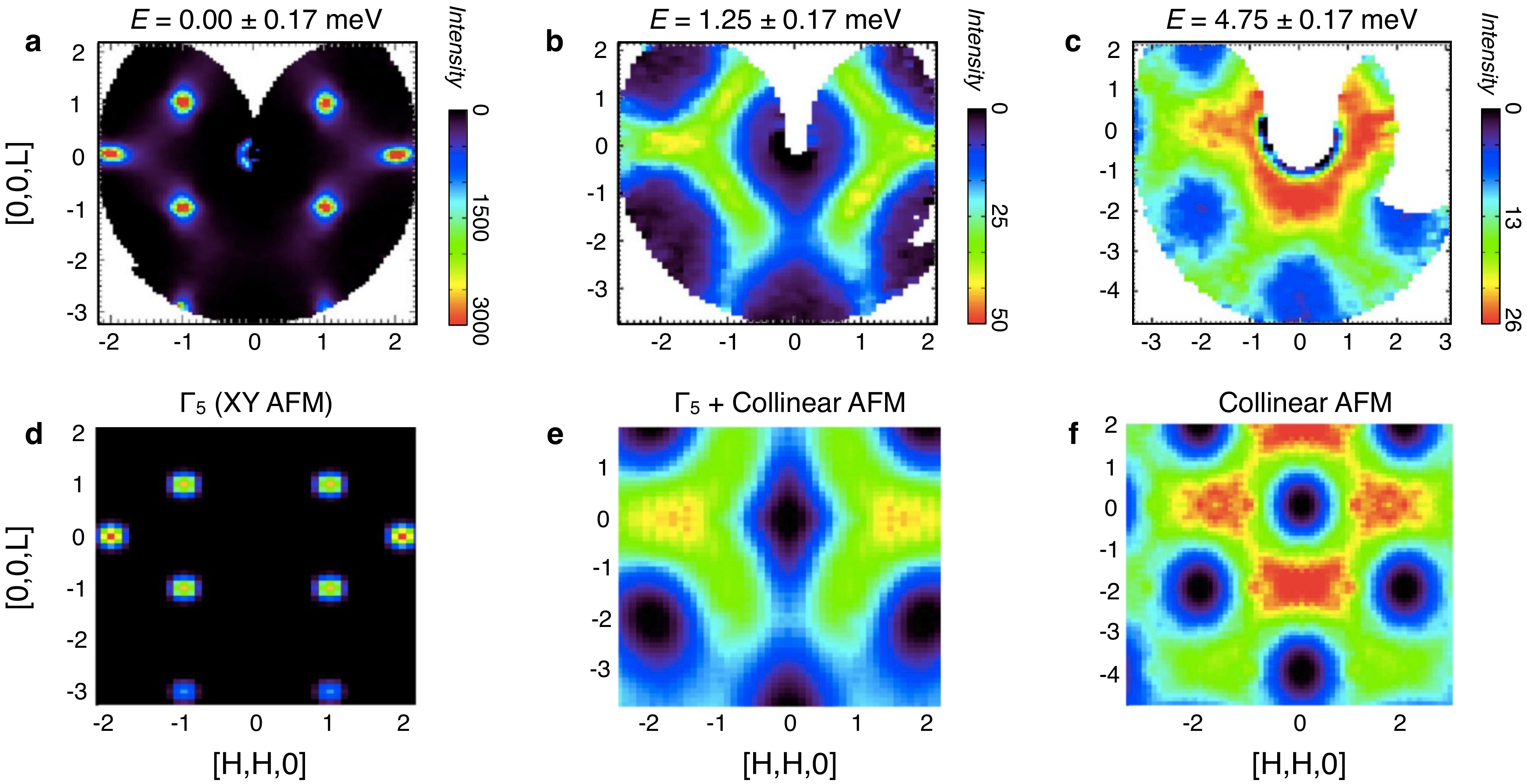}
\caption{The top row shows inelastic neutron scattering maps of NaCaCo$_2$F$_7$ at various energy transfers, (a) Elastic, $\Delta E = 0.00$~meV, (b) $\Delta E = 1.25$~meV, and (c) $\Delta E = 4.75$~meV, all with an integration of $\pm0.17$~meV. The bottom row shows calculated spectra that are consistent with the experimental data, indicating static short range order in the XY $\Gamma_5$ manifold and excitations consistent with both $\Gamma_5$ and a collinear antiferromagnetic state. Adapted from Ref.~\cite{ross2016static} with permission, copyrighted by the American Physical Society. }
\label{Fluoride_Pyrochlores}
\end{figure}

While detailed investigations of this new family have only recently begun, already there have been interesting developments for XY phenomenology with NaCaCo$_2$F$_7$, where the magnetism originates from Co$^{2+}$ in an effective $S=\frac{1}{2}$ state~\cite{ross2017single}. This material undergoes a spin freezing transition at $T_f = 2.4$~K due to weak exchange disorder arising from the random distribution of Na$^+$ and Ca$^{2+}$ on the $A$ sublattice \cite{krizan2014nacaco}. Elastic neutron scattering measurements and simulations for NaCaCo$_2$F$_7$, shown in \textbf{Figure~\ref{Fluoride_Pyrochlores}(a)} and \textbf{(d)}, reveal that within this frozen state, clusters of spins achieve short range order within the XY ($\Gamma_5$) manifold \cite{ross2016static}, and are argued to form a mosaic of $\psi_2$ and $\psi_3$~\cite{sarkar2016unconventional}. However, the inelastic spectra at energy transfers of $\Delta E = 1.25$~meV is described by a combination of XY and collinear antiferromagnetic spin configurations, as shown in \textbf{Figure~\ref{Fluoride_Pyrochlores}(b)} and \textbf{(e)}. At still higher energy transfers, $\Delta E = 4.75 $~meV, the spectra can be described by a collinear antiferromagnetic model alone (\textbf{Figure~\ref{Fluoride_Pyrochlores}(c)} and \textbf{(f)}). Ultimately, the spin freezing transition in NaCaCo$_2$F$_7$ preempts the breaking of the $\Gamma_5$ degeneracy and the formation of a long range ordered state~\cite{ross2016static}. However, the coexistence of dynamics from the XY ($\Gamma_5$) manifold and those originating from a collinear antiferromagnet indicate that phase competition may also be a factor in this material. 

In this review we have focused our attention on the cubic pyrochlore lattice. However, other exciting avenues exist to investigate XY anisotropy with related crystalline architectures. One such class of materials is the rare earth chalcogenide spinels, Cd$R_2$Se$_4$ and Cd$R_2$S$_4$, where the rare earth cation $R$ again forms a 3D network of corner-sharing tetrahedra~\cite{lau2005geometrical}. In a reversal from the case of the rare earth pyrochlores, erbium in this geometry has local Ising anisotropy, resulting in a spin ice state \cite{lago2010cder}. However, it has been predicted that a dysprosium variant, CdDy$_2$Se$_4$, would realize XY spin anisotropy \cite{wong2013ground}, making it a good candidate to explore the types of phenomenology discussed here, possibly including phase competition. A little further afield, another direction for future work is the so-called ``tripod kagome'' systems, $R_3$Mg$_2$Sb$_3$O$_{14}$ and $R_3$Zn$_2$Sb$_3$O$_{14}$, where $R$ is a rare earth cation \cite{dun2016magnetic,sanders2016synthesis,dun2017structural}. This chemical formula can be thought of in terms of a doubling of the pyrochlore lattice, where a quarter of the rare earth sites are replaced by antimony. The resulting structure is a two-dimensional kagome lattice of rare earth cations. Detailed investigations of these materials have only recently begun. However, the ytterbium and erbium variants of this family may retain the XY anisotropy of the pyrochlore parent compounds, and in this regard would make interesting case studies.

\begin{marginnote}[]
\entry{Kagome Lattice}{A two-dimensional lattice made up of corner sharing triangles and a canonically frustrated architecture.}
\end{marginnote}

\section{SUMMARY AND FUTURE ISSUES}

\begin{summary}[SUMMARY POINTS]
\begin{enumerate}
\item The nature of their crystal electric field ground states imparts the erbium and ytterbium pyrochlores with XY spin anisotropy and effective $S=\frac{1}{2}$. 
\item Many of the observed properties of the ytterbium and erbium pyrochlores can be attributed to the combination of anisotropic exchange and XY spin anisotropy. The general form of the nearest neighbor anisotropic exchange Hamiltonian produces a rich phase diagram, and competition between phases is likely key to understanding their exotic low temperature properties.
\item The experimental signatures of phase competition are multiple heat capacity anomalies, suppressed $T_N$ or $T_C$, and anomalous spin dynamics.
\item Many of the peculiarities of ground state selection in Yb$_2$Ti$_2$O$_7$, including extreme sample and pressure dependence, can be qualitatively understood within the scenario of phase competition.
\item Er$_2$Ti$_2$O$_7$, with a single relatively high temperature heat capacity anomaly, appears insensitive to phase competition and correspondingly, is robust to weak disorder.

\end{enumerate}
\end{summary}

\begin{issues}[FUTURE ISSUES]
\begin{enumerate}
\item Can phase competition within the anisotropic exchange model quantitatively account for all the exotic features observed in the XY pyrochlores, including their multiple heat capacity anomalies?
\item What is the origin of the unconventional spin excitations observed in the ytterbium XY pyrochlores?
\item With the prediction of emergent quantum electrodynamics, the physical properties of quantum spin ice are remarkable. Are any of these emergent properties, such as photon-like excitations, manifest in the XY pyrochlores?
\item Could strong phase competition result in a quantum spin liquid state, and what would the experimental signatures of this spin liquid be?
\item Can order-by-disorder be uniquely distinguished as the ground state selection mechanism in Er$_2$Ti$_2$O$_7$?
\end{enumerate}
\end{issues}

\section*{DISCLOSURE STATEMENT}
The authors are not aware of any affiliations, memberships, funding, or financial holdings that might be perceived as affecting the objectivity of this review. 

\section*{ACKNOWLEDGMENTS}
The authors appreciate helpful conversations with L. Balents, O. Benton, M.J.P. Gingras, E. Kermarrec, G.M. Luke, T.M. McQueen, R. Moessner, J.G. Rau, K.A. Ross, N. Shannon, C.R. Wiebe, M.N. Wilson, H. Yan, and M.E. Zhitormisky. This work was supported by NSERC of Canada.

\bibliographystyle{arstyle4}
\bibliography{Review_References2}

\end{document}